\documentclass[aps,pra,twocolumn,amsmath,amssymb,floats,floatfix,showpacs]{revtex4}

\usepackage{amsmath}              
\usepackage{graphicx}
\usepackage{dcolumn}

\begin{document}

\title{Decoherence control in different environments}   
\author{J. Paavola}
\email[]{janika.paavola@utu.fi}
\homepage[]{www.openq.fi}

\affiliation{
Turku Centre for Quantum Physics, Department of Physics and Astronomy, University of Turku,
FI-20014 Turun yliopisto, Finland
}

\author{S. Maniscalco}
\email[]{sabrina.maniscalco@utu.fi}
\homepage[]{www.openq.fi}

\affiliation{
Turku Centre for Quantum Physics, Department of Physics and Astronomy, University of Turku,
FI-20014 Turun yliopisto, Finland
}

\date{\today}    

\begin{abstract}
We investigate two techniques for controlling decoherence, focusing on the crucial role played by the environmental spectrum. We show how environments with different spectra lead to very different dynamical behaviours. Our study clearly proves that such differences must be taken into account when designing decoherence control schemes.
The two techniques we consider are reservoir engineering and quantum-Zeno control.
We focus on a quantum harmonic oscillator initially prepared in a nonclassical state and derive analytically its non-Markovian dynamics in presence of different bosonic thermal environments. On the one hand we show how, by modifying the spectrum of the environment, it is possible to prolong or reduce the life of a Schr\"odinger cat state. On the other hand we study the effect of nonselective energy measurements on the degradation of quantumness of initial Fock states. In this latter case we see that the crossover between Zeno (QZE) and anti-Zeno (AZE) effects, discussed by Maniscalco et al. [Phys. Rev. Lett. {\bf 97}, 130402 (2006)], is highly sensitive to the details of the spectrum. In particular, for certain types of spectra, even very small variations of the system frequency may cause a measurement-induced acceleration of decoherence rather than its inhibition.
\end{abstract}

\maketitle

\section{Introduction}\label{intro}

Environment induced decoherence, describing the transition from the quantum to the classical world, has been elaborately studied in the past decade (see, e.g., Refs. \cite{Zurek03,Decoherence}). Such a phenomenon is considered both a major obstacle in quantum information processing, and an ally in understanding the mechanisms underlying the quantum to classical transition \cite{Zurek03,Decoherence}. Generally, the unavoidable interaction of quantum systems with their environments is responsible for the transformation of a quantum state into a classical one. Such a process is faster and faster the more macroscopic the initial state is \cite{Strunz2003}. The details of the dynamics of the decoherence process depend, among other things, on the structure of the environment \cite{Paavola09}.

We study a quantum harmonic oscillator weakly coupled to a bosonic thermal bath. This model is one of the few ones, in the theory of open quantum systems, that is amenable to an exact solution \cite{breuer}. The system is a quantum harmonic oscillator, and therefore can be prepared in both highly quantum states, such as Fock states and Schr\"{o}dinger cat states, and semiclassical states such as coherent states and thermal states. Hence, it is very suitable for studying decoherence and loss of nonclassicality.

The dynamics of the reduced system obey the exact time-local Hu-Paz-Zhang master equation \cite{HuPazZhang}. The existence of an exact analytical solution of the Hu-Paz-Zhang master equation \cite{Ford01,analytic_solution} enables us to investigate the non-Markovian dynamics of the system. Since decoherence is a very rapid process the non-Markovian dynamics often play a crucial role. Moreover, the analytic non-Markovian solution is a key ingredient in the development of Zeno-control strategies since it makes it possible to trace back the origin of Zeno or anti-Zeno dynamics to the form of the reservoir spectrum.

Environment induced decoherence in quantum Brownian motion has been studied extensively in the past decades \cite{Strunz2003,Horhammer2008,Oh09,Zurek91}. Some studies consider also the effect different environments have on decoherence, focussing in particular on those cases in which the Markovian approximation, neglecting system-reservoir correlations, holds \cite{HuPazZhang,Paz93d,Villar07,Villar05,Buzek93,Buzek92}.

The study of the influence of different environments  on the open system dynamics, and in particular on the decoherence and loss of nonclassicality, is particularly timely. Experiments dealing with more and more complicated engineered reservoirs, from optical and microwave cavities \cite{Quang95,Quang97,John97,Haroche} to photonic crystals \cite{photonic_chrystal}, from controllable Ohmic environments \cite{Turchette00,Myatt00} to sub-Ohmic and super-Ohmic reservoirs \cite{Bollinger09}, are becoming, indeed, more and more accurate. The ability to modify in a controlled way the coherence properties of the system by acting on its environment, and in particular by modifying its spectral properties, necessitates, however, non-Markovian theoretical approaches since structured reservoirs are characterized by non-negligible memory effects.

In this paper we investigate how the quantum to classical transition can be modified by reservoir engineering for an initial Schr\"odinger cat state, and by nonselective energy measurements in the case of an initial Fock state. In the former case we compare three different Ohmic-like reservoirs and find out which one induces the slowest decoherence. In the latter case we show that the measurements may either slow down the quantum to classical transition, i.e., quantum Zeno effect, or speed up the transition, i.e., anti-Zeno effect (See \cite{Facchi08} and references therein for a review on QZE and AZE).

Quantum Zeno phenomena have been mostly studied using a two-level system model. In connection to quantum measurement theory the QZE was also discussed for a more complicated system and for different types of reservoirs in Ref. \cite{Shaji04}. Recently the description of QZE and AZE for the damped quantum harmonic oscillator has been given in the Ohmic reservoir case \cite{Maniscalco2006,Maniscalco10}. Here we generalize these results to the sub-Ohmic and super-Ohmic environment and bring to light the extreme sensitivity of these quantum phenomena to the form of the environment.

The paper is organized in the following way. Section II introduces our system, the non-Markovian master equation describing the dynamics, and the reservoir types we consider in the paper. In Sec. III we focus on reservoir engineering as a tool for changing the decoherence times, comparing the dynamics of an initial Schr\"odinger cat state for three different environments. In Sec. IV we show how to modify the quantum to classical transition by means of the quantum Zeno or anti-Zeno effect, for different reservoir spectra. Finally, in Sec. V we present the conclusions.

\section{The system}
The system we study is a quantum harmonic oscillator linearly coupled to a thermal reservoir modeled as an infinite chain of independent quantum harmonic oscillators. The total Hamiltonian is

\begin{equation}\label{eq:systemhamilton}
H=H_S+H_E+H_{int},
\end{equation}
where the Hamiltonians of the system oscillator, environment and interaction read

\begin{eqnarray}
             H_S &=& \omega_0\left(a^\dagger a+\frac{1}{2}\right), \\
             H_E &=& \sum_{n=0}^{\infty}\omega_n\left(b_n^\dagger b_n
+\frac{1}{2}\right), \\
             H_{int} &=& \frac{1}{\sqrt{2}}\,(a+a^\dagger)\sum_n k_n
(b_n+b_n^\dagger).
           \end{eqnarray}

As usual, $a \,(a^\dagger)$ and $b_n \,(b^\dagger_n)$ are the annihilation (creation) operators of the system and the environment oscillators, respectively, $\omega_0$ and $\omega_n$ are the frequencies of the system and the environment oscillators  and $k_n$ describes how strongly the system oscillator interacts with each mode of the reservoir. In the continuum limit one introduces the spectral density $J(\omega)$ defined as $J(\omega)=\sum_n k_n\delta(\omega-\omega_n)/(2m_n\omega_n)$, with $m_n$ the mass of the $n$-th environmental oscillator \cite{breuer}.
\subsection{Non-Markovian master equation}
Starting from the microscopic Hamiltonian \eqref{eq:systemhamilton}, an exact master equation can be derived for the reduced system. In the interaction picture this equation reads \cite{HuPazZhang, analytic_solution,rwa}
\begin{align}\label{ME}
\frac{d\rho(t)}{dt}=&-\Delta(t)[X,[X,\rho(t)]]\\\nonumber
&+\Pi(t)[X,[P,\rho(t)]]+\frac{i}{2}r(t)[X^2,\rho(t)]\\
&-i\gamma(t)[X,\{P,\rho(t)\}],\nonumber
\end{align}
where $\rho(t)$ is the reduced density matrix for the system oscillator, $X=(a+a^\dagger)/\sqrt{2}$ and $P=i(a^\dagger+a)/\sqrt{2}$. The coefficients $\Delta(t)$ and $\Pi(t)$ are the normal and anomalous diffusion coefficients, $\gamma(t)$ is the dissipation coefficient and $r(t)$ gives the time-dependent frequency shift \cite{HuPazZhang}.

The master equation (\ref{ME}) is exact, and therefore non-Markovian. The reservoir memory effects are encoded in the time-dependent coefficients. We note in passing that  time-local master equations are equivalent to master equations containing a memory kernel, in the sense that the latter ones can always be recast in time-local form \cite{Kossakowski10}.

In the weak coupling and high temperature regime $r(t)$ and $\Pi(t)$ can be neglected \cite{Intravaia03a}. In this case, and for times $t\ll t_{th}$, with $t_{th}$ the thermalization time, the approximate master equation describing the system dynamics is given by \cite{Maniscalco09}
\begin{align}\label{aME}
&\frac{d\rho(t)}{dt}=-\frac{\Delta(t)+\gamma(t)}{2}[2a\rho(t)a^\dagger-a^\dagger a\rho(t)-\rho(t)a^\dagger a]\\\nonumber
&+\frac{\Delta(t)-\gamma(t)}{2}[2a^\dagger\rho(t)a-a a^\dagger \rho(t)-\rho(t)a a^\dagger ]\\\nonumber
&+\frac{\Delta(t)-\gamma(t)}{2}e^{-2i\omega_0t}[2a\rho(t)a-a^2\rho(t)-\rho(t)a^2]\\\nonumber
&+\frac{\Delta(t)-\gamma(t)}{2}e^{2i\omega_0t}[2a^\dagger\rho(t)a^\dagger-(a^\dagger)^2\rho(t)-\rho(t)(a^\dagger)^2].
\end{align}
Decoherence occurs at time scales much shorter than $t_{th}$. Therefore the use of this master equation is justified throughout the paper.

In the secular approximation we coarse grain over time scales of the order of  $1/ \omega_0$ and therefore the last two terms of the Eq. \eqref{aME} average out to zero. The secular approximated master equation reads
\begin{align}\label{secME}
&\frac{d\rho(t)}{dt}=-\frac{\Delta(t)+\gamma(t)}{2}[2a\rho(t)a^\dagger-a^\dagger a\rho(t)-\rho(t)a^\dagger a]\\\nonumber
&+\frac{\Delta(t)-\gamma(t)}{2}[2a^\dagger\rho(t)a-a a^\dagger \rho(t)-\rho(t)a a^\dagger ].
\end{align}
We will further discuss the validity of the secular approximation  in Sec. \ref{reservoir}.
The diffusion and dissipation coefficients, in second order perturbation theory, take the form
\begin{align}
\label{eq:delta}
\Delta(t) =& 2\int_0^{t}dt'\,
\int_0^{\infty}d\omega\,J(\omega)\left[N(\omega)+\frac{1}{2}\right]\\
&\times\cos(\omega\nonumber
t')\cos(\omega_0 t'), \\
\gamma(t) =& 2\int_0^{t}dt'\,
\int_0^{\infty}d\omega\,\frac{J(\omega)}{2}\sin(\omega
t')\sin(\omega_0 t'),\label{eq:pikkugamma}
\end{align}
where $N(\omega)=(e^{ \omega/k_B T}-1)^{-1}$ is the average number of reservoir thermal excitations, $k_B$ the Boltzmann constant, and $T$ the
reservoir temperature. In the long time limit $t\gg t_{th}$ these coefficients attain their positive Markovian values $\Delta_M$ and $\gamma_M$, given by
\begin{align}
\Delta_M=&\pi I(\omega_0)\label{deltamarko}\\
\gamma_M=&\frac{\pi}{2}J(\omega_0).
\end{align}
In the next section we introduce and discuss the family of reservoir spectral densities used in the paper.

\subsection{Modeling the reservoir}\label{reservoir}
We consider reservoir spectral densities of the form
\begin{equation}\label{spectra}
J(\omega)=g^2\omega_{c}^{1-s}\omega^s e^{-\omega/\omega_c},
\end{equation}
where $s$ is a real parameter, $\omega_c$ is the cutoff frequency and $g$ a dimensionless coupling constant. We consider as examples  reservoirs with $s=1$, $3$ and $1/2$ corresponding to Ohmic, super-Ohmic and sub-Ohmic spectral densities, respectively. These types of reservoirs have been recently engineered in the trapped ion context \cite{Bollinger09} and a theoretical comparative study of the heating function for these reservoirs types has been presented in Ref. \cite{Paavola09}.

The spectral distribution
\begin{equation}
I(\omega)=J(\omega)\left[N(\omega)+\frac{1}{2}\right]
\end{equation}
contains all the necessary information about the environment. In this paper we focus on the high temperature regime where $I(\omega) \simeq J(\omega) k_B T/\omega$. The spectral distributions of the reservoirs under study are shown in Fig. 1.

\begin{figure}\label{fig:reservoirs}
\includegraphics[width=7cm]{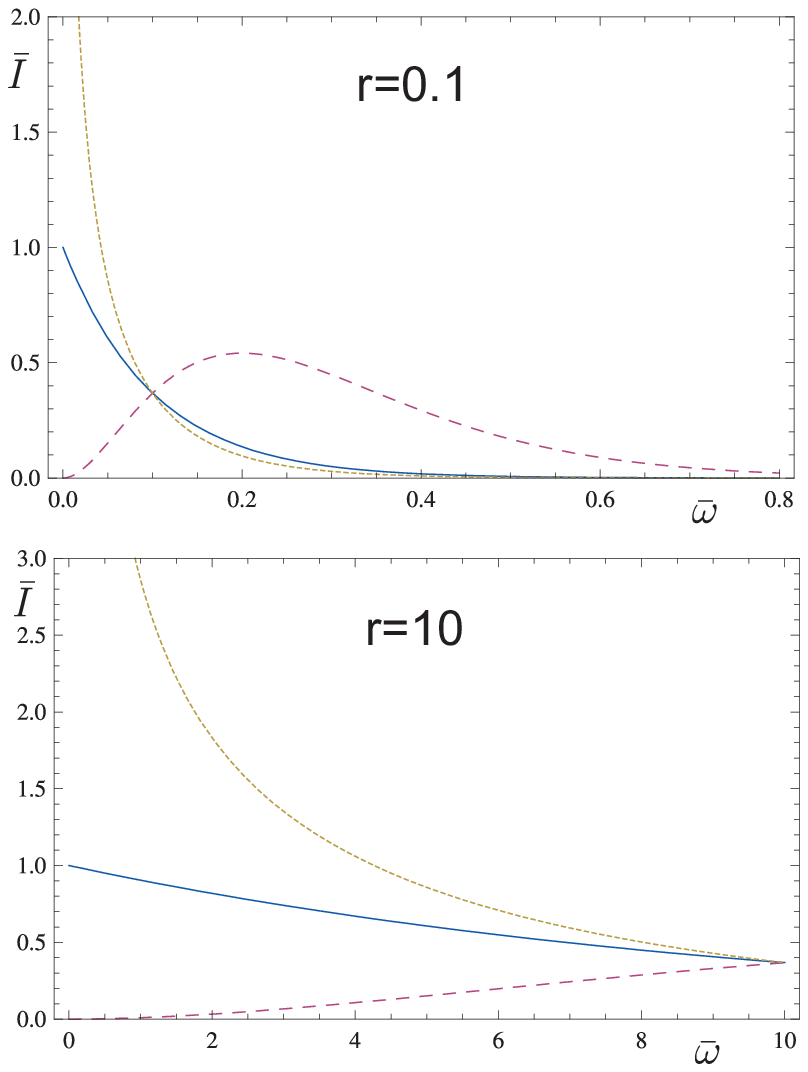}
\caption{Spectral distributions $\bar{I}=I/(g^2 k_BT)$ for the Ohmic (solid), super-Ohmic (dashed) and sub-Ohmic (dotted line) high-$T$ reservoirs, vs $\bar{\omega}=\omega/\omega_0$. In the upper figure we plot the off-resonant $r=0.1$ case, while the lower figure shows the $r=10$ case. For each spectral curve the location of the cutoff frequency is given by $\bar{\omega}_c=\omega_c/\omega_0 = r$. The location of the system oscillator frequency is always $1$.}
\end{figure}

A crucial role in the environment induced dynamics of the system  is played by the resonance parameter, defined as the ratio between the cutoff frequency and the system oscillator frequency,
\begin{equation}
r=\frac{\omega_c}{\omega_0}.
\end{equation}
Changing this parameter corresponds to shifting the system oscillator frequency with respect to the reservoir spectrum. This allows to control the effective coupling between the system and the environment \cite{Myatt00,Turchette00}. For $r\ll1$ the system oscillator is off-resonant with respect to the peak of the reservoir spectrum for all the three reservoir types we consider.

From previous results \cite{Paavola09, Maniscalco04,Maniscalco09} we expect to see different types of dynamics in the $r\ll1$ and $r\gg 1$ regimes.
Since we are interested in the non-Markovian dynamics occurring at time scales $\omega_c t\le 1$, we can use the secular master equation \eqref{secME} when $r\ll1$, since in this case the secular approximation holds in the non-Markovian time scales. For $r\gg 1$, on the other hand, the secular approximation cannot be performed at short times $\omega_ct\le 1$. Therefore, in this latter case, we must use the master equation \eqref{aME}.

In the following section we will define the tool used to characterize decoherence of a Schr\"odinger cat state, namely the fringe visibility function, we will present the analytic solutions of both master equations \eqref{aME} and \eqref{secME} and we will examine how decoherence occurs, within these two parameter regimes, for different engineered reservoirs.

\section{Controlling decoherence via reservoir engineering}\label{Sec:reseng}

Let us consider as initial state a Schr\"odinger cat state of the form
\begin{equation}\label{cat}
|\Psi\rangle=\frac{1}{\sqrt{\mathcal{N}}}(|\alpha\rangle+|-\alpha\rangle),
\end{equation}
where $|\alpha\rangle$ is a coherent state and   $\mathcal{N}^{-1}=2[1+\mathrm{exp}(-2|\alpha|)^2]$. For simplicity we assume $\alpha$ real. The Wigner function $W(\beta)$, with $\beta \in \mathbb{C}$, for this state consists of two Gaussian peaks centered in $\beta=(\alpha,0)$ and $\beta=(-\alpha,0)$, and an interference term in between the peaks. The interference term signals the quantumness of the superposition and it is absent for classical statistical mixtures. The disappearance of the interference term is thus considered a mark of the quantum to classical transition.

To follow the dynamics of the decoherence process, it is convenient to look at the fringe visibility function \cite{Paz93d}
\begin{align}
F(\alpha,t)&\equiv\mathrm{exp}(-A_{int})\nonumber\\
&=\frac{1}{2}\frac{W_I(\beta,t)|_{\mathrm{peak}}}{[W^{(+\alpha)}(\beta,t)|_{\mathrm{peak}}W^{(-\alpha)}(\beta,t)|_{\mathrm{peak}}]^{1/2}}\label{fringevis},
\end{align}
where $W_I(\beta,t)|_{\mathrm{peak}}$ is the value of the Wigner function at $\beta=(0,0)$ and $W^{(\pm\alpha)}(\beta,t)|_{\mathrm{peak}}$ are the values of the Wigner function at $\beta=(\pm\alpha,0)$, respectively.
Our aim is to study the dynamics for the sub-Ohmic, Ohmic and super-Ohmic reservoirs in order to identify the form of the spectrum leading to the slowest environment induced decoherence.
We consider separately the cases $r\gg 1$ and $r\ll 1$.

\subsection{The off-resonant case $r\ll1$}
The solution of Eq. \eqref{ME} in terms of the quantum characteristic function was derived in Ref. \cite{analytic_solution}. The corresponding Wigner function is written as the sum of three terms: two describing the evolution of the peaks and one giving the interference term dynamics \cite{Maniscalco09}. For the initial state here considered, one gets, for $r\ll1$,
\begin{equation}\label{fringevis01}
F(\alpha,t)=\mathrm{exp}\left[-2\alpha^2\left(1-\frac{e^{-\Gamma(t)}}{2N(t)+1}\right)\right],
\end{equation}
where
\begin{align}
N(t) =& \int_0^{t}dt'\,
\Delta(t'), \\
\Gamma(t) =& 2\int_0^{t}dt'\,
\gamma(t').
\end{align}
%

\begin{figure}\label{fig:decoherence_offreso}
\includegraphics[width=7cm]{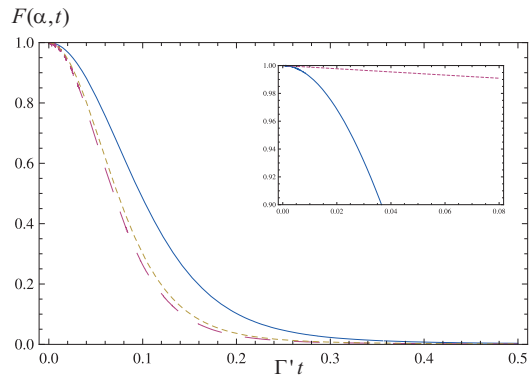}
\caption{Fringe visibility for $r=\omega_c/\omega_0=0.1$ for the Ohmic (solid), super-Ohmic (dashed) and sub-Ohmic (dotted line) reservoirs. The inset shows the comparison between non-Markovian (dotted line) and Markovian (solid line) fringe visibility for the Ohmic reservoir. Plots are given in unitless time $\Gamma't=2g^2\omega_0t$, where $g$ is the coupling constant. We have set $k_B T/(\hbar \omega_0)=100$.}
\end{figure}
\subsection{The resonant case $r\gg1$}
In the opposite regime, i.e., $r\gg1$, the fringe visibility function reads \cite{Maniscalco09}
\begin{equation}\label{fringevis10}
F(\alpha,t)=\mathrm{exp}\left[-2\alpha^2\left(1-\frac{e^{-\Gamma(t)}}{4N(t)+1}\right)\right].
\end{equation}
Note that the only difference between the fringe visibility function in the off-resonant ($r\ll 1$) and resonant ($r\gg 1$) regimes is  a factor of two appearing in front of the mean energy of the oscillator $N(t)$, also known as the heating function, for $r\gg 1$. This means that the heating process, in the $r\gg 1$ case, can be seen as due to an effective reservoir at a temperature $2T$. This difference stems from the fact that, in the resonant regime we do not neglect the counter rotating terms in Eq. \eqref{aME}. These terms provide two additional channels for the energy exchange between the system and the environment. This is consistent with what was found for the heating function in Ref. \cite{EPJB03}.

\subsection{Similarities and differences in the dynamics}

The time evolution of the fringe visibility for the Ohmic, sub-Ohmic and super-Ohmic reservoirs is shown in Figs. 2 and 3, for $r \ll 1$ and $r \gg 1$, respectively. The initial non-Markovian quadratic behavior, as opposed to the exponential one typical of flat Markovian reservoirs, is clearly visible. As an example we have plotted the Markovian versus non-Markovian fringe visibilities in the insets of Figs. 2 and 3 for the Ohmic reservoir and for short initial times. We note that in the off-resonant case, see Fig. 2, the Markovian fringe visibility decays slower than the non-Markovian one. This is due to the initial jolt in $\Delta(t)$ causing a faster decoherence for $r\ll 1$. On the contrary, in the resonant case, see Fig. 3, the Markovian decay of the fringe visibility is faster than the non-Markovian one. In the latter case, indeed, $\Delta(t)<\Delta_M$, hence the initial non-Markovian decoherence is slower. The other reservoirs show similar behavior in the Markovian vs non-Markovian initial dynamics.

The time evolution of the fringe visibility factor shows a similar qualitative behavior for all three reservoir types (Ohmic, sub-Ohmic and super-Ohmic), both in the resonant and in the off-resonant regime.  This is in contrast with the dynamics of the heating function where non-Markovian oscillations, indicating an exchange of energy between the system and the environment, characterize the  $ r \ll 1 $ regime, for all types of reservoirs \cite{Paavola09}.

In general, the decoherence process is significantly faster for $r\gg 1$ than for $r\ll 1$. Indeed, in the former case the effective coupling of the system to the reservoir is stronger than in the off-resonant case, due to the overlap between the frequency of the system oscillator and the reservoir spectrum. Moreover, in the $r\gg 1$ case the system interacts with the engineered reservoir via two effective channels, due to the non-negligible effect of the counter-rotating terms, as explained in the previous subsection.

The Ohmic reservoir induces the slowest decoherence,  while the super-Ohmic and sub-Ohmic reservoirs decay in a very similar manner, both faster than 
the Ohmic case. Therefore, if one is able to modify the natural reservoir spectrum into an Ohmic form,
 one would slow down decoherence with respect to the sub-Ohmic and super-Ohmic ones, and in the case of $r=10$ also with respect to the corresponding Markovian reservoir.

Note that,  for $t\ll t_{th}$, the exponential factors in Eqs. (\ref{fringevis01}) and (\ref{fringevis10}) can be approximated to one. This tells us that decoherence depends essentially only on the diffusion coefficient $\Delta(t)$ through the heating function $N(t)$. The interaction with the reservoir causes both decoherence and heating/dissipation. For the system studied in this paper, and for $t\ll t_{th}$, these two processes are both characterized by the same coefficient $\Delta(t)$. It is straightforward to check that, in the decoherence time scale, the heating of the system is very small. Therefore, there exists a clear distinction between the decoherence and heating time scales. However one can see, e.g., in the $r=10$ case,  that the heating induced by an Ohmic reservoir is much slower, than the one caused by a super-Ohmic or sub-Ohmic reservoir. This fact also justifies the slowest decoherence experienced by the system in the Ohmic case, as shown in Fig. 3.
\begin{figure}\label{fig:decoherence_reso}
\includegraphics[width=7cm]{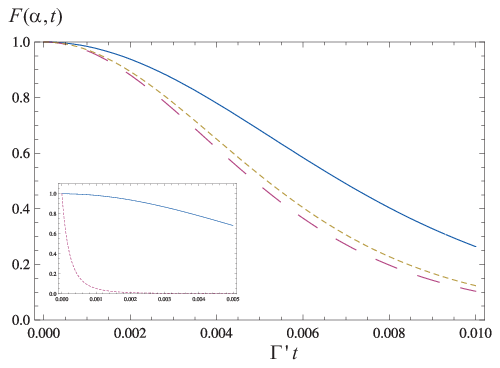}
\caption{Fringe visibility for $r=\omega_c/\omega_0=10$ for the Ohmic (solid), super-Ohmic (dashed) and sub-Ohmic (dotted line) reservoirs.  The inset shows the comparison between non-Markovian (dotted line) and Markovian (solid line) fringe visibility for the Ohmic reservoir. Plots are given in unitless time $\Gamma't=2g^2\omega_0t$, where $g$ is the coupling constant. We have set $k_B T/(\hbar \omega_0)=100$.}
\end{figure}
%
\begin{figure*}\label{fig:zeno}
\includegraphics[width=18cm]{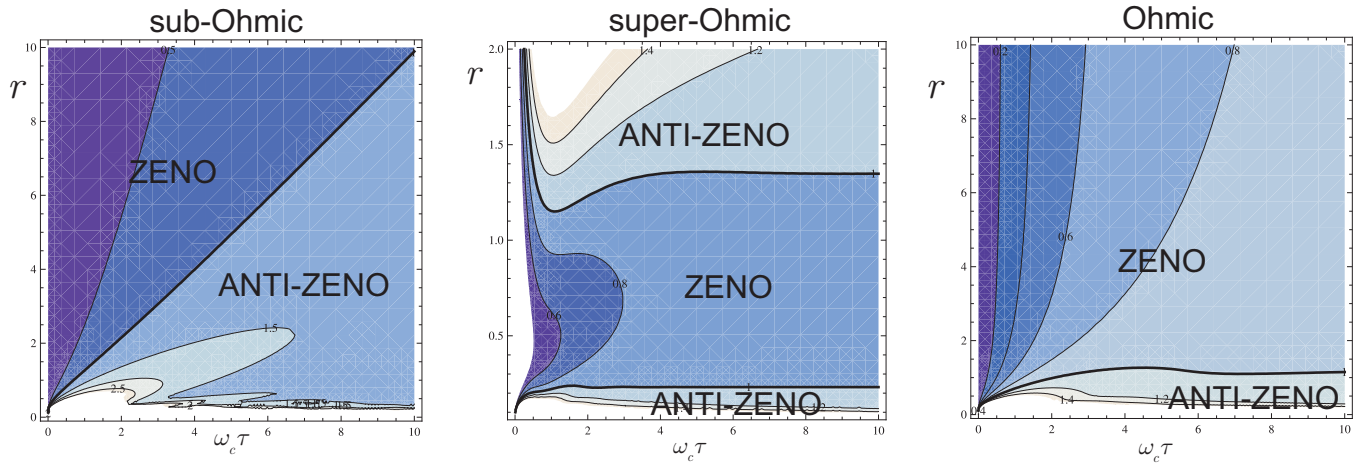}
\caption{QZE/AZE -crossover for the sub-Ohmic, super-Ohmic and Ohmic high $T$ reservoirs. We plot the ratio between the effective decay rate $\gamma^Z$ and the Markovian decay rate $\gamma^0$ as a function of $r=\omega_c/\omega_0$ and of the measurement time interval $\omega_c \tau$.}
\end{figure*}
In the next section we will investigate another way of modifying the transition from the quantum to the classical world based on a completely different approach, i.e.,  by performing frequent measurements on the system.

\section{Controlling decoherence via the Quantum Zeno effect}
It is known that the decay of an unstable system can be altered by making frequent measurements on the system \cite{Pascazio01}. For the quantum Brownian motion case here studied, it has been shown that, in the Ohmic case, nonselective energy measurements performed in the short non-Markovian timescale may either inhibit or enhance the quantum to classical transition, as a consequence of the quantum Zeno or anti-Zeno effect, respectively \cite{Maniscalco2006}.

In this section we aim at exploring the crossover between Zeno and anti-Zeno dynamics for the three different reservoirs introduced in Sec. \ref{reservoir}. Elucidating the role played by the reservoir spectrum in the occurrence of the Zeno or anti-Zeno effect is important due to the fact that different physical realizations of, e.g., a qubit have different reservoir spectra. For example, it is known that solid state qubits are subjected to sub-Ohmic noise ($1/f$ noise) while optical qubits normally interact with an almost flat Markovian spectrum.

As we will see in the following, the crossover between Zeno and anti-Zeno dynamics is extremely sensitive to the details of the spectrum, especially in the super-Ohmic case. Moreover, contrarily to the decoherence control strategy based on reservoir engineering presented in Sec. \ref{Sec:reseng}, in the case of measurement-based control  different reservoir spectra lead to very different dynamics.

We consider an initial Fock state $|n\rangle$. We assume to perform  $N$ non-selective energy measurements at time intervals $\tau$ such that $P_n(\tau)= \langle n \vert \rho_S(\tau) \vert n \rangle \simeq 1$. The survival probability, i.e., the probability that the system is still in its initial state $|n\rangle$ at time $t=\tau N$ is then given by \cite{Facchi2001,Maniscalco2006}
\begin{equation}
P_n^N (t)=P_n(\tau)^N\equiv\mathrm{exp}[-\gamma_n^Z(\tau)t].
\end{equation}
Here $\gamma_n^Z(\tau)$ is the effective decay rate. At high temperatures this rate is given by $\int_0^\tau\Delta(t') \,dt'/\tau$ and it is independent of $n$ \cite{Maniscalco2006}. Let us denote with $\gamma^0$ the decay rate of the survival probability in the absence of measurements. This quantity corresponds to the limit $\tau\rightarrow\infty$, i.e.,
\begin{equation}
\gamma^0=\lim_{\tau\rightarrow\infty} \gamma_n^Z(\tau)=\Delta_M,
\end{equation}
with $\Delta_M$ given by Eq. \eqref{deltamarko}. We note that both $\gamma_n^Z$ and $\gamma^0$ depend on the reservoir spectrum.

The crossover between Zeno and anti-Zeno dynamics is given by \cite{Maniscalco2006}
\begin{equation}\label{eq:QZE-AZE}
\frac{\gamma_n^Z(\tau)}{\gamma_n^0}\simeq\frac{\int_0^\tau\Delta(t') \,dt'}{\tau\Delta_M}.
\end{equation}
If a finite time $\tau^*$ such that $\gamma_n^Z(\tau^*)=\gamma^0$ exists, then, for times $\tau<\tau^*$, we have $\gamma_n^Z(\tau)/\gamma^0<1$, i.e., the decay in presence of measurements is slower than the Markovian decay in absence of measurements (QZE). On the other hand, for $\tau>\tau^*$, $\gamma_n^Z(\tau)/\gamma^0>1$ and an acceleration of the decay due to the measurements occurs (AZE).

In Fig. 4 we show a contour plot of Eq. (\ref{eq:QZE-AZE}), for the three different reservoirs under consideration, as a function of the parameter $r$ and of the interval between the measurements $\omega_c \tau$.  The QZE/AZE -crossover is indicated by a bold solid contour line. Note that the time $\tau^*$ identifying the crossover strongly depends on $r$. In particular, both in the super-Ohmic and in the Ohmic case, for some value of $r$, only the QZE occurs, and the time $\tau^*$ does not exist. This is in contrast with the AZE-dominated dynamics of radiative decay described in Ref. \cite{Kurizki00}. On the other hand, for both the super-Ohmic and the Ohmic spectra, there exist also values of $r$ in correspondence of which two $\tau^*$ exist, as, e.g., the value $r=1.2$ in the super-Ohmic case.  For this value of $r$ one can see from Fig. 4 that, increasing $\tau$, one passes from Zeno to anti-Zeno dynamics and then again to Zeno dynamics.

For the Ohmic reservoir the AZE occurs only for $r<1$. Therefore, for  Ohmic environments with $r>1$, measurements can only prolong the life of quantum states as the initial Fock state here considered. For the sub-Ohmic reservoir, on the other hand, there exists always a cross-over between Zeno and anti-Zeno dynamics, for any value of the resonance parameter $r$. In particular, by increasing the measurement interval $\tau$ one passes from a situation in which decoherence is slowed down to a situation in which it is enhanced.

The super-Ohmic reservoir presents some additional interesting features. The dynamics is mostly anti-Zeno dominated, except, of course, for very small values of $\tau$.  The two AZE regions are disconnected by a narrow band of QZE region between $0.2\lesssim r\lesssim1.3$. This situation indicates that, for a given super-Ohmic spectrum, two system oscillators with slightly different frequencies $\omega_0\simeq \omega_c$ (correspondent, e.g., to $r=1$ and $r=1.5$)  may act, in presence of measurements, in completely opposite ways, one showing mostly AZE ($r=1.5$) and the other one only QZE ($r=1$). The occurrence of this type of behavior gives a clear indication of the sensitive role played by the system and reservoir parameters.

A common feature shared by all the reservoirs is that, for $r \ll 1$,  nonselective energy measurements always accelerate decoherence. The reason lies in the initial jolt of the diffusion coefficient $\Delta(t)$, which causes an initial decoherence much stronger than in the Markovian case  \cite{Maniscalco2006}. The off-resonant regime is also characterized by strong non-Markovian features  such as oscillations and regions of negativity in the diffusion and dissipation coefficients. However, the AZE occurs also when the time-dependent coefficients are positive, e.g., in the super-Ohmic case for $r>1$. Also in this case an initial jolt  is present in $\Delta(t)$  (see Fig. 4.1 in Ref. \cite{PaavolaThesis}).

\section{Conclusions}
In this paper we have compared two different strategies for controlling environment-induced decoherence for a quantum harmonic oscillator interacting with a high-$T$ bosonic bath. The first strategy is based on reservoir engineering, a new technique that has been demonstrated recently in many physical context, e.g., in trapped ion systems \cite{Bollinger09}. We have seen that an initial Schr\"{o}dinger cat state is transformed in the corresponding statistical mixture more slowly in an Ohmic engineered reservoir than in a super-Ohmic or sub-Ohmic reservoir.

It is worth noticing that the quantum to classical transition indicated by the disappearance of the interference fringes in the Wigner function never presents strongly non-Markovian features, such as oscillations, when the time-dependent coefficients oscillate attaining negative values. This behavior is different from the case of a two state system in a coherent superposition of two orthogonal states. For a two state system, indeed, the non-Markovian quantum jumps approach shows that the occurrence of temporarily negative decay rates can be interpreted in terms of reverse quantum jumps restoring the quantum superposition destroyed by a previously occurred quantum jump \cite{nmqj,nmqj2}. This situation never occurs for the  Schr\"{o}dinger cat dynamics here considered. Indeed, even in the $r \ll 1$ regime, the coherence between the two coherent states forming the superposition is never partly restored.

The second technique for controlling decoherence is based on the QZE and AZE. We study how sensitive these effects are to the form of the natural reservoir spectrum. Our results on the crossover between the QZE and the AZE show that some types of environment are more sensitive than others to the reservoir parameters. The super-Ohmic reservoir, for example, shows a remarkable sensitivity to the value of the parameter $r$. Indeed, slight changes in $r$ may give rise, for the same value of $\tau$, to either the QZE or the AZE.

The quantum Zeno effect is known to be closely connected to decoherence control methods \cite{Facchi05}. The very rich variety of Zeno and anti-Zeno dynamics for this system makes it extremely interesting for testing fundamental features of quantum physics such as the possibility of controlling the quantum to classical transition by means of energy measurements \cite{Facchi05}. In view of the astonishing advances in both the coherent manipulation of single quantum systems and the reservoir engineering techniques, we believe that this phenomenon will be soon in the grasp of the experimentalists.
\acknowledgments

Financial support from the Turku Collegium of Science and Medicine (S.M.), the Emil Aaltonen foundation, the Finnish Cultural foundation and the V\"ais\"al\"a foundation is gratefully acknowledged.

\end{document}